# Nonlinear Interference in Crystal Superlattices

Anna V. Paterova, and Leonid A. Krivitsky*

*Institute of Material Research and Engineering (IMRE),*
*Agency for Science Technology and Research (A*STAR), 138634 Singapore*
*E-mail: Leonid_Krivitskiy@imre.a-star.edu.sg

Nonlinear interferometers with correlated photons hold a promise to advance optical characterization and metrology techniques by improving their performance and affordability. Nonlinear interferometers offer the sub-shot noise phase sensitivity and enable measurements in detection-challenging regions using inexpensive and efficient components. The sensitivity of such interferometers, defined by the ability to measure small shifts of interference fringes, can be significantly enhanced by using multiple nonlinear elements, or crystal superlattices. However, to date, the experiments with no more than two nonlinear elements have been realized, thus hindering the potential of nonlinear interferometers. Here, we build a nonlinear interferometer with up to five nonlinear elements in a highly-stable and versatile configuration. We study the modification of the interference pattern in different configurations of the superlattice and perform a proof-of-concept gas sensing experiment with enhanced sensitivity. Our approach offers a viable path towards broader adoption of nonlinear interferometers with correlated photons for imaging, interferometry, and spectroscopy.

## Introduction

Optical characterization and metrology techniques benefit from using correlated photons, particularly in studies of light-sensitive and fragile biological and chemical samples [1-3]. For example, strong temporal correlations between photons were used for a single-photon calibration of the efficiency of retinal cells [4] and enhancing the nonlinear response of biological samples [5]. Furthermore, two-photon interference effects formed the basis for the dispersion-free optical coherence tomography [6-8], microscopy with enhanced phase contrast [9], noise-robust spectroscopy of nanostructures [10] to name a few.

Recently, the nonlinear interference of correlated photons attracted particular interest in the context of infrared (IR) metrology and sensing [11-15]. The nonlinear interferometer is composed of two nonlinear elements, which produce pairs of correlated photons (signal and idler) under coherent excitation. The signal (in the visible range) and idler (in the IR range) photons are mixed in the interferometric setup, and as long as one cannot distinguish which nonlinear element produced the photons, the interference fringes are observed. The interference pattern for signal photons depends on the phases and amplitudes of the signal, idler, and pump photons. When idler photons interact with a sample, its properties in the IR range can be inferred from the interference pattern of signal photons in the visible range. Thus the technique addresses practical challenges of generation and detection of the IR light since the sample response is obtained using accessible components for visible light.

Nonlinear interferometers have been realized using numerous physical platforms, including bulk nonlinear crystals [11-14, 16-18], gas cells [19], fiberized networks [20, 21], and nonlinear waveguides [22, 23]. Nonlinear interferometers were used for imaging [24], spectroscopy [16, 25-27], optical coherence tomography [28, 29], super-resolution interferometry [18, 19] and polarimetry [30]. All these techniques are intrinsically interferometric. Hence their sensitivity is defined by the ability to detect small changes in the interference pattern, such as the shift of the fringes or change of the fringe visibility.

One possible way of enhancing the sensitivity of nonlinear interferometers was outlined by D. Klyshko [31], who considered a setup with *N* identical nonlinear elements separated by linear gaps, referred here as a *crystal superlattice*. He showed that with the increase of the number of crystals, bright interference fringes narrow down, yet the spacing between fringes remains unchanged. The idea was theoretically expanded in more recent works [20, 32, 33] however, to the best of our knowledge, there are no reports on the experimental realization of nonlinear interferometers with more than two nonlinear elements. The major challenges in its practical realization are associated with (1) the necessity of superimposing signal and idler modes from multiple nonlinear elements while preserving the quantum indistinguishability, and (2) the necessity to align and stabilize increasingly complex setup.

Here, for the first time, we realize a nonlinear interferometer with a crystal superlattice consisting of up to five nonlinear elements. In our setup, nonlinear elements are arranged sequentially and are pumped by a single coherent laser. By careful design and alignment, we achieve a robust mode overlap of signal and idler photons with remarkable stability. We observe the interference pattern in the frequency-angular spectrum with full flexibility of crystal arrangements. We also perform a proof-of-concept gas sensing experiment with enhanced sensitivity.

**Theory**

Let us consider *N* identical nonlinear crystals of length *l* separated by *N-1* equal linear gaps *l'*, see Fig. 1. The crystals are pumped by a coherent laser, and each crystal produces signal (*s*) and idler (*i*) photons via spontaneous parametric down-conversion (SPDC). Down-converted photons from each crystal are re-directed to the next crystal. The state of the two-photon field from a single crystal is given by [34]:

$$|\psi\rangle = |vac\rangle + \sum_n \sum_{k_s, k_i} f_n(\vec{k}_s, \vec{k}_i) a^+_{k_{ns}} a^+_{k_{ni}} |vac\rangle, \qquad (1)$$

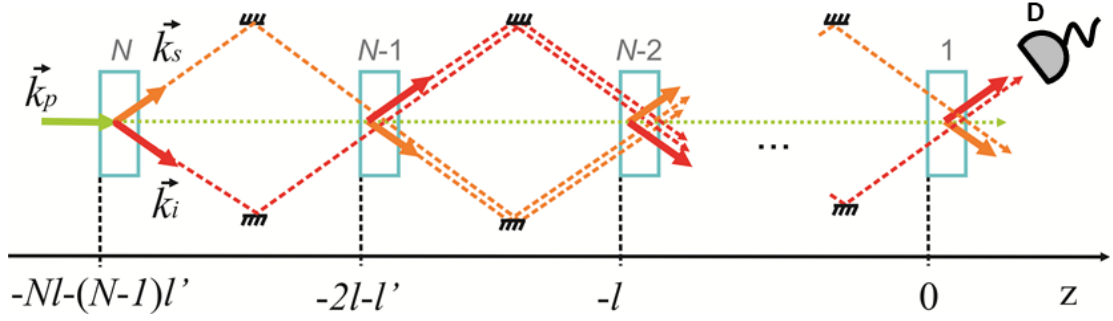

*Figure 1. Schematics of the nonlinear interferometer with crystal superlattice. N identical nonlinear crystals, separated by equal gaps, are coherently pumped by a laser (green arrow). In each crystal, the pump photon $k_p$ decays into a pair of correlated signal $k_s$ (orange arrow) and idler $k_i$ (red arrow) photons, which are then redirected to the next crystal. The intensity of signal photons is measured in the experiment by the detector D. The paths of signal and idler photons are disjoined for clarity.*

where $f_n(\vec{k}_s, \vec{k}_i)$ is the two-photon field amplitude from the $n$-th crystal $n = [1, N]$, $a^+_{k_{ns}}$ and $a^+_{k_{ni}}$ are creation operators of photons in $n$-th crystal with wavevectors $\vec{k}_s$ and $\vec{k}_i$ respectively, and $|vac\rangle$ is the vacuum state.

Assuming the pump is a monochromatic plane wave, and the crystal is thin and uniform, the amplitude of a two-photon field is given by [31, 35]:

$$f_n \propto \chi E_p \int_{z_n}^{z_{n+1}} dz\, D_n^{*p} D_n^s D_n^i, \qquad (2)$$

where $\chi$ is the second-order susceptibility of the crystal; $z_n = -nl + (n-1)l'$ is the coordinate of the front edge of the $n$-th nonlinear crystal, $E_p$ is the field of the pump, $D_n^j$ is the propagation function for signal, idler and pump photons ($j = s, i, p$):

$$D_n^j(k_j, z) = \exp\left[-ik_j^z z + (n-1)(k_j'^z - k_j^z)l'\right], \qquad (3)$$

where $k_j^z$ and $k_j'^z$ are the longitudinal wave vectors inside the nonlinear crystal and in the gap between crystals, respectively. From Eqs. (2) and (3) we obtain the two-photon field amplitude:

$$f_n = \frac{1 - \exp(-i\Delta kl)}{i\Delta kl} \cdot \left[-i(n-1)(\Delta kl + \Delta k'l')\right], \qquad (4)$$

where $n = [1, N]$, $\Delta k$ and $\Delta k'$ are the wave vector mismatches inside the nonlinear crystal and in the linear gap, respectively. For $N$ identical crystals, the two-photon field amplitude is given by the sum of contributions from individual crystals:

$$F = \sum_{n=1}^{N} f_n \propto \mathrm{sinc}(\Delta kl/2) \sum_{n=1}^{N} e^{i(n-1)\varphi}, \qquad (5)$$

where $\varphi = (\Delta kl + \Delta k'l')$. Then, from Eq. (5) the intensity distribution of the signal photons, measured in the experiment, is given by [31, 36]:

$$I_N \propto \left\{ \text{sinc}\left(\frac{\Delta k l}{2}\right) \cdot \frac{\sin[N\varphi/2]}{\sin[\varphi/2]} \right\}^2. \tag{6}$$

We express the phase mismatch $\Delta k$ in frequency and scattering angle, following [37, 38], and plot interference patterns for the nonlinear interferometer with two and five crystals, see Fig. 2(a, b), respectively. Fig. 2(c) shows cross-sections of the interference patterns for different number of crystals in the superlattice. We see that as the number of crystals increases, the interference maxima become narrower yet the spacing between them remains unchanged.

In Section 1 of Supplementary Materials, we show that the width of the fringes is inversely proportional to the number of crystals $N$:

$$\delta\theta_s \propto \frac{\pi}{N}. \tag{7}$$

From Eq. (7) we note the striking similarity between the interference fringes in the nonlinear interferometer with the crystal superlattice and a conventional multi-slit linear interference [39].

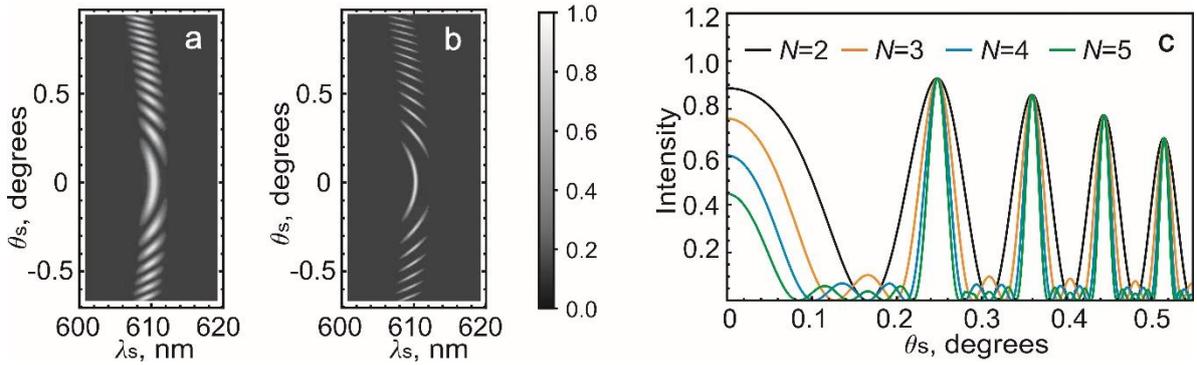

*Figure 2.(a, b) Calculated frequency-angular spectra (interference patterns) of signal photons from two (a) and five (b) identical $LiNbO_3$ crystals (crystal length l=1mm, air gaps between crystals l'=8.2 mm, orientation of the axis $\theta_c$ =50.34°, the pump is a 532 nm laser). (c) Vertical cross-sections of (a, b) at $\lambda_s$=610.4 nm (idler photon wavelength is $\lambda_i$=4142 nm) for a different number of crystals in the superlattice.*

**Experiment**

Our experimental setup is shown in Fig.3. A continuous-wave laser (cw) with wavelength 532 nm (60 mW, Laser Quantum) pumps a set of identical Lithium Niobate nonlinear crystals cut from a single master crystal (5%MgO:$LiNbO_3$, $l$ = 1 mm, cut angle of 48.5°, Eksma Optics). The crystals are separated by the distance $l'$ = 8.2 mm. Photon pairs are generated in each nonlinear crystal in type-I quasi-collinear frequency non-degenerate regime. A notch filter NF and a polarizer V are used to filter out the pump. Signal photons are focused on the slit of the imaging spectrograph (Acton) using the lens F ($f$=300 mm). The interference pattern of signal photons in frequency-angular coordinates is recorded by a CCD camera for visible light (Andor iXon 897).

To ensure the indistinguishability of photon pairs produced in every crystal of the superlattice, all the SPDC photons should be generated and propagate within the interaction volume defined by the pump beam. This requirement is expressed in the following condition $(2l + l')\tan(\theta_s) \ll d$, which links the scattering angle $\theta_s$, pump diameter $d$, and parameters of the superlattice $l, l'$ : [16, 26]. To satisfy this condition, we set $d$ ~3 mm using the beam expander (LS) and detect angles up to $\theta_s=\pm 0.85°$.

Obtaining interference patterns with high visibility requires careful alignment of the interferometer. First, the orientation of each crystal is set to generate identical frequency spectra, which are measured by the spectrograph, see Section 2 of Supplementary materials. Then, by observing pairwise interference fringes between crystals, we ensure that the optical axes of the crystals are aligned in the same direction. Next, distances between the crystals are carefully aligned to ensure equal gaps between them, see Section 3 of Supplementary materials. Each crystal is mounted on a 2D translation stage so that it can be moved in and out of the interferometer. By successively observing interference patterns from two, three, and four crystals, we adjust the distances between the crystals such that the fringes are overlapped. The accuracy of setting the length of the gap $l'$ between crystals by this method is better than 100 μm. After the alignment of the crystals, we perform measurements of the interference with different numbers of crystals in the superlattice.

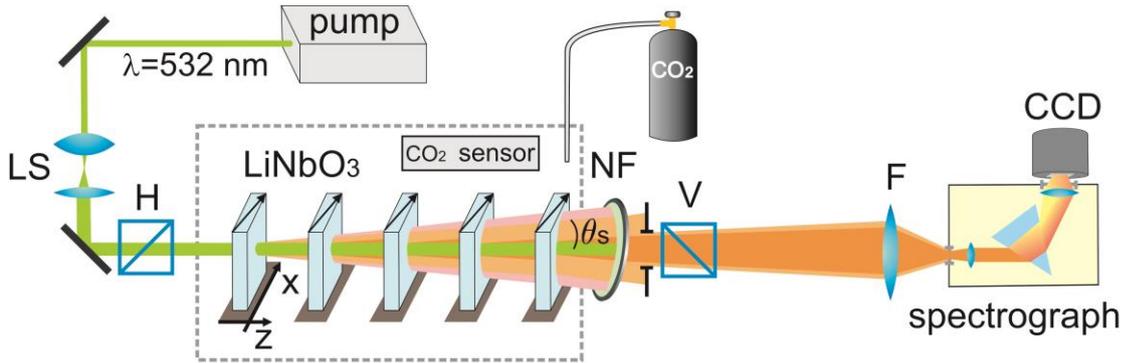

*Figure 3. Experimental realization of the nonlinear interferometer with crystal superlattice. A cw-laser pumps SPDC crystals. Signal photons are shown in orange, and idler photons are shown in pink. Photons generated in different crystals are overlapped within the interaction volume, defined by the pump beam. The signal photons are projected by the lens F onto the slit of the imaging spectrograph equipped with the CCD camera. The crystals can be moved in and out of the optical path; their optical axes are aligned in the same direction (marked by the arrows). In the gas sensing experiment the carbon dioxide gas is injected into the enclosure (marked by a dashed rectangle).*

In the gas sensing experiment, the interferometer is placed in an airtight enclosure with the input socket for the carbon dioxide gas ($CO_2$, 99.9% purity). The wavelength of the idler photons is set to match the absorption peak at around 4.27 μm. The gas homogeneously fills up the volume between the crystals. Its concentration in the enclosure is controlled by a commercial $CO_2$ sensor (Amphenol).

## Results and discussion

**Observation of the interference with the crystal superlattice.** First, we set the phase-matching angle at $\theta_c=50.34°\pm0.02°$, when the signal SPDC photons are generated around 610.4 nm (bandwidth 2 nm) and idler photons at 4.14 μm (bandwidth 92 nm). The normalized frequency-angular spectra of signal photons for two and five nonlinear crystals are shown in Figs. 4(a, b), respectively. Our key observation is that the interference fringes for the interferometer with five crystals get narrower comparing to the interferometer with two crystals, yet the period of the fringes remains unchanged. Fig. 4(c, d) shows the cross-sections of the interference pattern at $\lambda_s=610.4$ nm for the interferometer with two and five crystals, respectively. Cross-sections are taken by averaging the intensity across the bandwidth of $\Delta\lambda_s=0.4$ nm. The acquisition time is 200 seconds for each measurement.

Solid curves in Figs. 4(c, d) correspond to theoretical calculations. The green curve shows the theory in the ideal case, and the red curve shows the theory which accounts for the experimental accuracy in setting the phase-matching angle of each crystal $\Delta\theta_c=\pm0.02°$. We found that the slight misalignment in setting the $\theta_c$ becomes crucial for the interferometer with the increasing number of crystals. The thorough analysis of the sensitivity of the interferometer to various experimental parameters is presented in Sections 4 and 5 of Supplementary Materials.

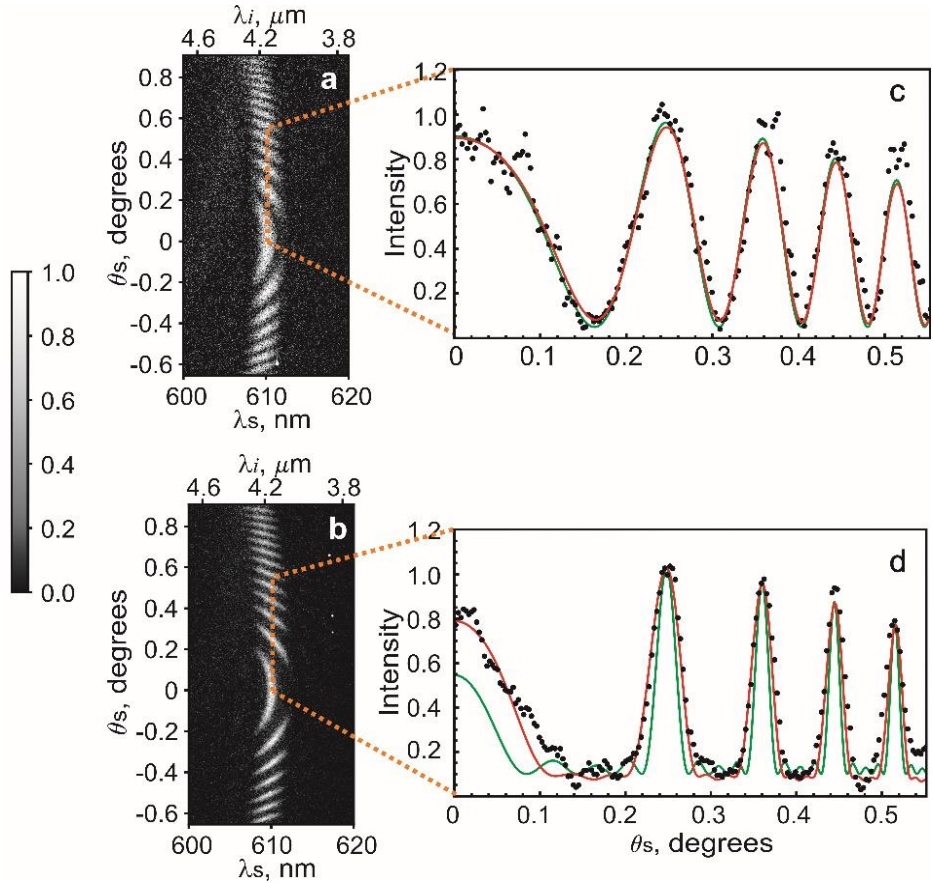

*Figure 4. (a, b) Frequency-angular spectra (interference patterns) of signal photons for the nonlinear interferometer with (a) two and (b) five crystals. The bottom abscissa axis shows the detected*

wavelength of signal photons $\lambda_s$, while the top abscissa axis indicates the wavelength of conjugated idler photons $\lambda_i$. The grayscale for both graphs shows the intensity normalized to the maximum value in each experiment. (c, d) Cross-sections of the interference fringes at $\lambda_s=610.4$ nm for an interferometer with two (c) and five (d) crystals. Black dots are experimental data, and solid lines are theoretical calculations. The green curve in (c, d) shows calculations for the ideal case, and the red curve shows calculations, taking into account the uncertainty in setting the phase-matching angle of each crystal by $\Delta\theta_c=\pm0.02°$.

We experimented with sets of two, three, four, and five crystals in the superlattice. In each case, we fit the experimental data with Eq.(6) and determine the width of the interference fringes. Fig. 5 shows the ratio of the widths of the interference fringes for the *N*-crystal interferometer $\delta\theta_{sN}$ and two-crystal interferometer $\delta\theta_{s2}$. The key observation is consistent with the theory: the interference fringes become narrower with increasing the number *N* of nonlinear crystals. The green dots in Fig.5 show linear scaling of the relative width in the ideal case, see Eq.(7). Red dots in Fig.5 show calculation results taking into account the uncertainty in the setting of the phase-matching angle of each crystal $\Delta\theta_c=\pm0.02°$, which is consistent with our experimental data, shown by black squares. Note that stronger dependence on the uncertainty of experimental parameters in the interferometer with crystal superlattice is a manifestation of the common property of multi-element interferometers.

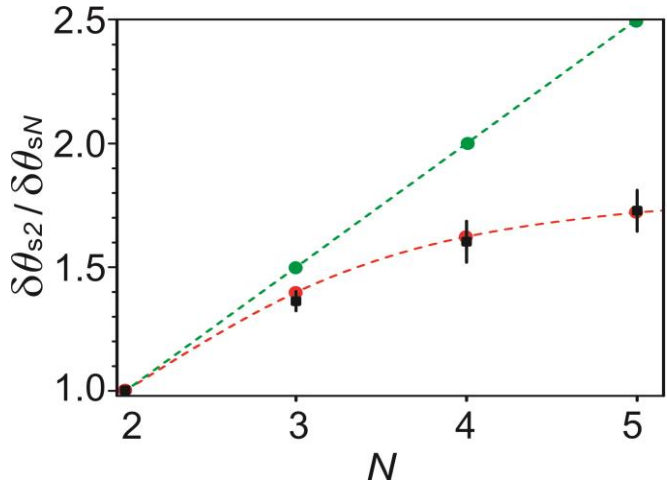

*Figure 5. Experimental dependence of the ratio of the width of fringes $\delta\theta_{s2}/\delta\theta_{sN}$ on the number of crystals in the superlattice (black squares). The green dots show theoretical calculations in the ideal case; the red dots show the calculated dependence which accounts for the experimental uncertainty in setting the phase-matching angle $\Delta\theta_c=\pm0.02°$. Dashed lines are given to guide the eye.*

**A proof-of-concept gas sensing experiment.** We set the phase-matching angle at $\theta_c=50.1°\pm0.02°$ and obtain signal photons at $\lambda_s=608.3$ nm and idler photons in the vicinity of the absorption resonance of the $CO_2$ at $\lambda_i=4.24$ μm. The normalized frequency-angular spectra of signal photons of an interferometer with two and five crystals are shown in Figs. 6(a, b) and 6(c, d), respectively. Fig. 6 (a, c) corresponds to the case when there is air in the gap between the crystals, and Fig.6(b, d) corresponds to the case when $CO_2$ gas is injected in the gaps (concentration 0.02%). Because of the absorption of idler photons by the gas, the interference pattern of signal photons experiences the phase shift and reduction of visibility.

Fig.7 shows the cross-sections of the interference pattern at $\lambda_s$=607 nm ($\lambda_i$=4.3 μm), when the wavelength of idler photons is detuned from the absorption resonance of $CO_2$ for about 40 nm. In this case, the gas causes a phase shift of the interference fringes without significant change of the fringe visibility. Fig. 7(a, b) correspond to interference fringes in an interferometer with two and five crystals, respectively. The reference measurement is taken with the air between the crystals.

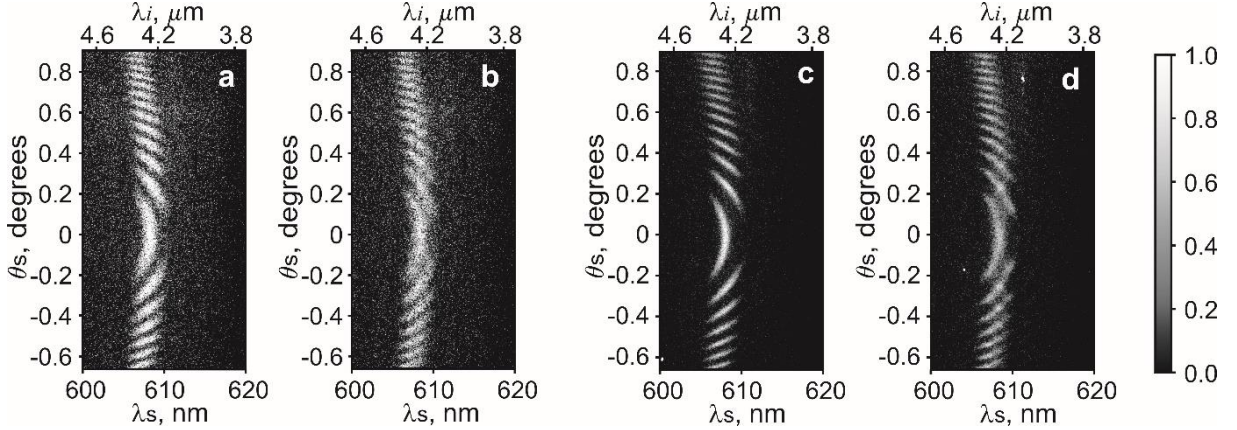

*Figure 6. The normalized frequency-angular spectra (interference patterns) for the interferometer with two (a, b) and five (c, d) crystals. (a) and (c) show the reference interference pattern with the air gap between crystals, (b) and (d) correspond to the data when $CO_2$ gas is injected.*

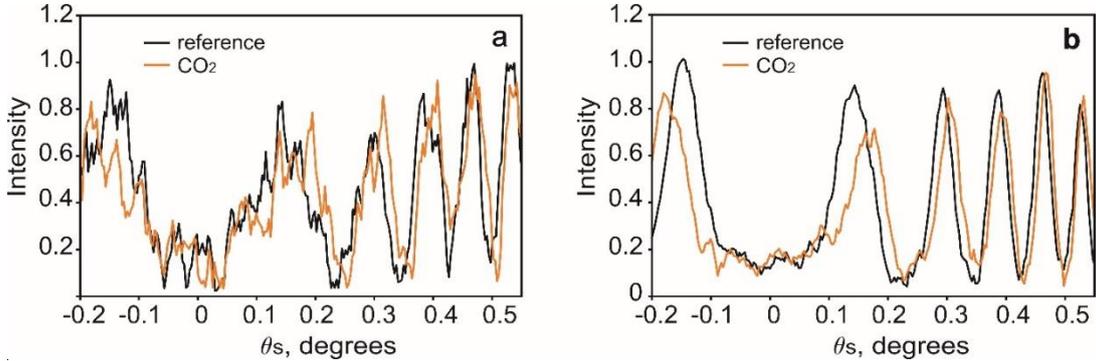

*Figure 7. Cross-section of the interference patterns at $\lambda_s$=607 nm for nonlinear interferometer with (a) two and (b) five crystals. The black curve shows the reference interference fringes with the air gap between crystals. Orange curves show interference fringes with the injected $CO_2$ gas.*

In both cases, the shift interference fringes relative to the reference is about $\Delta\varphi \approx 0.23\pm0.02$ π. However, in the five-crystal interferometer, the shift is more obvious due to the increase of the slope of the interference fringes by a factor of 1.6±0.2. Furthermore, in the five-crystal interferometer, the signal-to-noise ratio significantly improves.

**An interferometer with a "defect" in the superlattice**. Our experimental setup allows full flexibility in investigating nonlinear interferometers with variable crystal configurations. To demonstrate this, we remove the third crystal from the interferometer and observe the interference from the first, the second, the fourth, and the fifth crystal (assuming air between the crystals). We used Eq.(5) to calculate the interference pattern, which in this case, is given by:

$$I \propto \left\{ \text{sinc}\left(\frac{\Delta k l}{2}\right) \cdot \cos\left[\Delta k l + \frac{3\Delta k' l'}{2} + \frac{\Delta k' l}{2}\right] \cos\left[\frac{\Delta k l}{2} + \frac{\Delta k' l'}{2}\right] \right\}^2. \qquad (8)$$

The theoretical interference pattern, given by Eq.(8), is shown in Fig.8(a), and the corresponding experimental result is shown in Fig.8(b). The results are found to be in good agreement. As one can see, the interference pattern contains additional contributions originating from interferometers with different gaps. The ability to manipulate the interference patterns opens up possibilities for quantum state engineering [20].

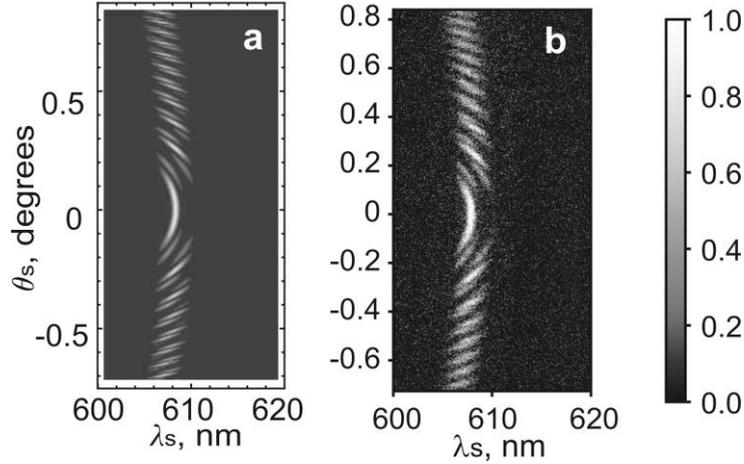

*Figure 8. Normalized frequency-angular spectra (interference patterns) in the case when the third crystal is removed from the interferometer. The interference from the first, second, fourth, and fifth crystal is observed: (a) theoretical and (b) experimental dependencies. The parameters of the interferometer are the same as in Figs.6, 7 with an air gap between the crystals.*

## Conclusions

We realized the nonlinear interferometer with crystal superlattice with up to five nonlinear elements. We experimentally demonstrated that with the increase of the number of nonlinear elements, the interference fringes become narrower, which directly translates to the improved sensitivity in metrological and sensing applications. The observed effect has a clear analogy with classical multi-source interference. We also found that the interferometer with crystal superlattice becomes increasingly dependent on the accuracy in setting experimental parameters, in particular, the phase-matching angles of the crystals. It reflects the common property of multiple-beam interferometers, which are more demanding to the settings of individual elements.

The presented configuration allows flexibility in the realization of unconventional crystal configurations for example by setting different gaps between crystals and using crystals of different sizes, which opens an interesting possibility for the quantum state engineering.

We anticipate that our work will trigger more than one creative design in the realization of complex nonlinear interferometers with correlated photons, such as by using mirrors, integrated photonics or fiber platforms. We believe that the presented concept will provide a viable path towards high-performance devices for sensing, metrology, and quantum state engineering.

## Acknowledgments

We acknowledge the support of the Quantum Technology for Engineering (QTE) program of A*STAR. We are grateful to Sergei Kulik, Radim Fillip, Maria Chekhova, Galya Kitaeva, and Berthold-Georg Englert for stimulating discussions.

# Supplementary materials

## 1. Calculation of the width of the interference fringes

The maxima of the interference pattern in Eq. (6) are observed, when $\varphi_{max} = 2\pi m$, $m = [0, \pm 1, \pm 2 ... \pm \infty)$. The positions of the intensity minima are determined by the fast oscillating function and observed at $\varphi_{min} = 2\pi m'/N$, $m' = [\pm 1, \pm 2 ... \pm \infty) \cup m' \neq mN$. The width of the interference fringes is determined as the difference between positions, where interference reaches maxima and the closest minima ($m' = mN + 1$):

$$\delta\varphi_N = 2\pi/N. \tag{A1}$$

As the phase mismatches $\Delta k$ and $\Delta k'$ are proportional to the square of the detection angle $\theta_s$, Eq. (A1) can be expressed as the following:

$$\delta\varphi_N = (\Delta k l + \Delta k' l')_{min} - (\Delta k l + \Delta k' l')_{max} \propto \left(C\theta_s^2\right)_{max} - \left(C\theta_s^2\right)_{min}. \tag{A2}$$

Expanding Eq. (A2) and using Eq. (A1), we obtain the following:

$$\delta\varphi_N \propto \left(\theta_{s,max} - \theta_{s,min}\right)\left(\theta_{s,max} + \theta_{s,min}\right) = \delta\theta_s\left(\theta_{s,max} + \theta_{s,min}\right) \propto 2\pi/N, \tag{A3}$$

where $\delta\theta_s$ is the width of the interference fringes in angular coordinates. Hence, the width of the fringes is inversely proportional to the number of nonlinear crystals in the interferometer:

$$\delta\theta_s \propto \frac{\pi}{N\left(\theta_{s,max} + \theta_{s,min}\right)/2} = \frac{\pi}{N\langle\theta_s\rangle}, \tag{A4}$$

where $\langle\theta_s\rangle$ is an average of $\theta_{s,max}$ and $\theta_{s,min}$. Note that the width of the fringes decreases with the detection angle. This can be clearly seen from the measurements.

Form Eq. (A4) the ratio of the width of the interference fringes in two- and $N$- crystal interferometer is given by:

$$\frac{\delta\theta_{s2}}{\delta\theta_{sN}} \propto \frac{N\langle\theta_{sN}\rangle}{2\langle\theta_{s2}\rangle} \approx \frac{N}{2}, \tag{A5}$$

here where we assume that $\langle\theta_{s2}\rangle \approx \langle\theta_{sN}\rangle$.

## 2. Spectral alignment of crystals in the superlattice

The orientation of each crystal is set to generate identical frequency spectra, which are measured by the spectrograph. Each crystal is adjusted separately by moving in and out of the optical path. The measured spectra of individual crystals are shown in Fig. S1.

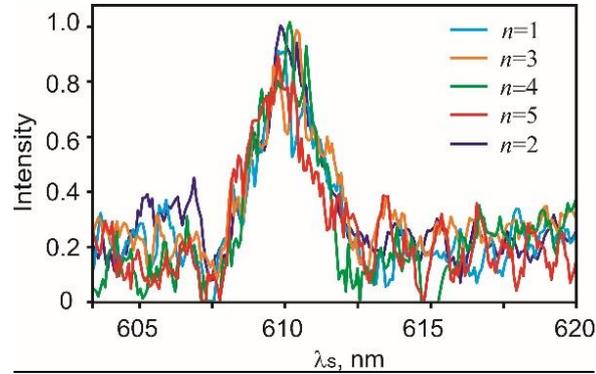

*Figure S1. The spectrum of signal photons from n-th crystal taken at $\theta_s=0°$. The orientation of each crystal is adjusted till spectrum for each crystal coincides.*

## 3. Alignment of the orientation and the gap between the crystals

Distances between the crystals are carefully aligned to ensure equal gaps between them. Each crystal can be moved in and out of the interferometer. By successively observing interference patterns from the interferometer with two, three, and four crystals, we adjust the distances between the crystals such that the fringes are overlapped, see Fig.S2.

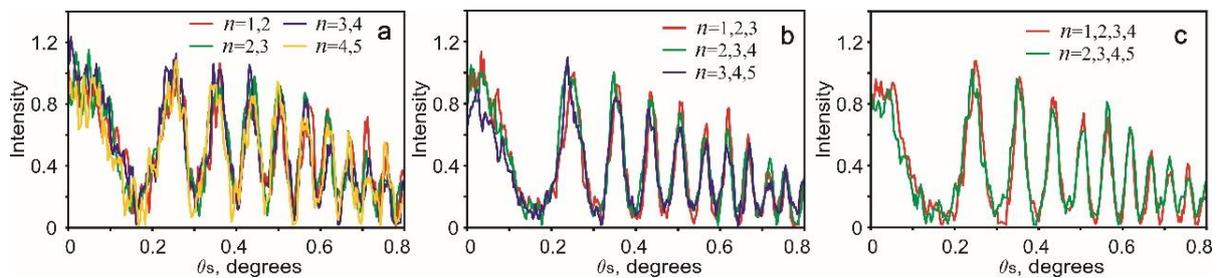

*Figure S2. Frequency-angular spectrum (Interference fringes) from different combinations of (a) two, (b) three, and (c) four nonlinear crystals. The distance between the crystals is adjusted so that the interference fringes are in phase.*

# 4. Sensitivity of the interference pattern to the uncertainty in the distance between the crystals.

We experimentally tested the sensitivity of the interference pattern for an interferometer with five crystals to a slight misalignment of the distance between the fourth and the fifth crystals. Misalignment of the gaps by 100 μm affects the visibility of the interference fringes only at larger angles, see Fig.S3.

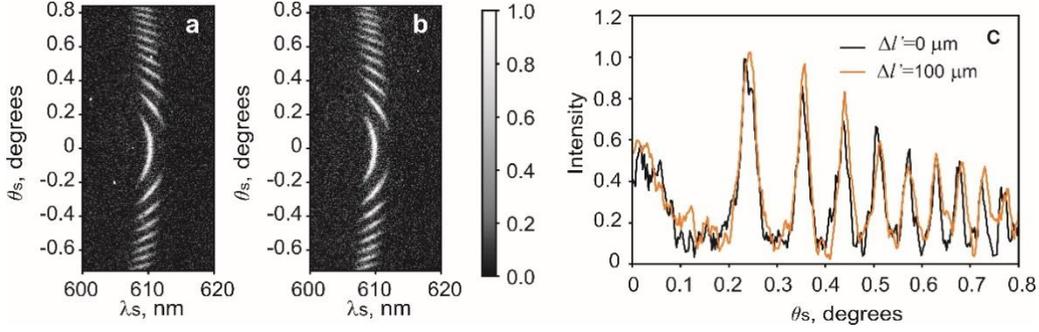

*Figure S3. Alignment of the gap between five crystals. (a) Interference pattern, when all gaps are aligned Δl'=0 μm. (b) Interference pattern, when the fifth crystal is misaligned by Δl'=100 μm. (c) shows cross-sections of (a) and (b) at Δλ$_s$=610 nm.*

# 5. Sensitivity of the interference pattern to the uncertainty in the crystal length, the gap width, and the orientation

We simulated the interference pattern for an interferometer with five crystals by varying the value of the crystal length $\Delta l$, the gap between crystals $\Delta l'=100$ μm, and the phase-matching angle θc. We found that the slight misalignment in setting of the phase-matching angle within $\Delta\theta_c=\pm 0.02°$, which corresponds to our experimental accuracy, becomes a key factor, which is responsible for the broadening of the interference fringes.

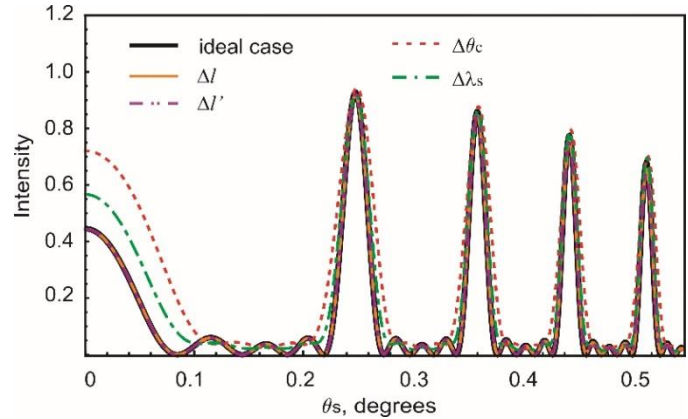

*Figure S4. Theoretical calculations of the interference fringes in the interferometer with five crystals for an ideal case (solid black curve) and for a small variation in crystal length Δl=±0.1 mm (solid orange curve), gap length Δl'=100 μm (dash-dotted purple curve), phase matching angle Δθ$_c$ =0.02° from each crystal (dashed red curve). Dash-dotted green curve is calculated taking into account averaging of the 5 data points along the wavelength Δλ$_s$=0.4 nm. The uncertainty in the setting of the phase-matching angle controls the interference pattern.*